# Quantum Fluctuations in Magnetically Frustrated $Gd_2GaSbO_7$ and $Gd_2InSbO_7$ Pyrochlores

Nallamuthu[1,*], Andrea Dzubinska[2,*], K. Arun[3], Vimaljith A. R.[1], R. Nagalakshmi[3], Ivan Čurlik[4], Sergej Iľkovič[4], Marian Reiffers[4]

[1]Department of Physics, School of Advanced Sciences, VIT-AP University, Amaravati, 522237, AP, India
[2]CPM-TIP, Pavol Jozef Safarik University in Kosice, Tr. SNP 1, 040 11 Kosice, Slovakia
[3]Department of Physics, National Institute of Technology, Tiruchirappalli 620015, Tamil Nadu, India
[4]Faculty of Humanities and Natural Sciences, Presov University, 080 01 Presov, Slovakia

[*]Email: nallamuthu.s@vitap.ac.in (S. Nallamuthu),
andrea.dzubinska@upjs.sk (Andrea Dzubinska)

## Abstract

$Gd_2GaSbO_7$ and $Gd_2InSbO_7$ pyrochlore compounds exhibit quantum fluctuations in a spin ice-like state. These compounds have not been adequately studied based on the concept of magnetic frustration. Here, we have synthesised and characterised $Gd_2GaSbO_7$ and $Gd_2InSbO_7$ to investigate their ground-state magnetic properties. We confirmed the cubic pyrochlore structure via X-ray powder diffraction. Magnetic and heat capacity measurements show an absence of magnetic order down to 400 mK. Interestingly, the low-temperature magnetic heat capacity of $Gd_2GaSbO_7$ and $Gd_2InSbO_7$ exhibits a broad maximum around $T_{max}$ = 0.9 K and $T_{max}$ = 1 K, respectively. The low-temperature $C_{mag}$ data fit with the power-law ($C_{mag} \approx T^{d/n}$) with dependencies of $T^{1.71}$ for $Gd_2GaSbO_7$ and $T^{1.49}$ for $Gd_2InSbO_7$, suggesting spin fluctuations caused by the disorder of magnetic moments at low temperatures. Both compounds retain residual entropy akin to water ice, resembling the spin ice state. Additionally, we analysed the ratio between nearest-neighbour exchange and dipolar interaction for all reported Gd-based pyrochlores, finding that $Gd_2InSbO_7$ has the lowest and weakest exchange interaction among them. This indicates that chemical pressure affects exchange interaction due to the different pathways of mediators between magnetic ions. This study reveals robust quantum fluctuations in the ground state in GGSO and GISO, similar to quantum spin ice compounds such as $Pr_2Zr_2O_7$, $Pr_2Sn_2O_7$, and $Yb_2GaSbO_7$. Other Gd-pyrochlore compounds exhibit long-range magnetic order, which is in stark contrast to this.

# Introduction

Geometrical frustration is one of the intriguing phenomena in magnetism. When magnetic spins are arranged in specific types of crystal lattices, such as triangular, pyrochlore, and Kagome lattices, they interact through competing exchange interactions with all the nearest-neighbouring spins. Consequently, no ordered magnetic state can be achieved. This leads to a large degeneracy of the ground state systems, where the spins are strongly correlated but still highly fluctuated at absolute zero temperature [1]. As a result, these systems exhibit a wide variety of interesting and unusual ground states, such as spin liquids and spin ice states[2-8]. The pyrochlore lattice, it is composed of corner-sharing tetrahedra. It can also be described as a three-dimensional structure with alternating stacks of Kagome and triangular layers, as shown in Figure 1. The general formula for pyrochlores is $A_2B_2O_7$, where both A and B form separate pyrochlore sublattices[9]. Many examples of pyrochlore materials have been reported, with rare earth cations occupying the magnetic A-site and non-magnetic cations occupying the B-site. Some members of the rare earth pyrochlore systems exhibit a variety of unusual magnetic ground states[4-7, 10]. One interesting behaviour is the Ising spin interaction of classical spin ice, as observed in $Dy_2Ti_2O_7$ [11-13], $Dy_2Sn_2O_7$ [14], $Ho_2Ti_2O_7$ [14-17], $Ho_2Sn_2O_7$ [14], and $Yb_2Ti_2O_7$ [10, 18, 19]. Spin ices consist of regular corner-sharing tetrahedra of magnetic ions, as shown in Figure 1 of the pyrochlore lattice. In these structures, the spins are constrained to point radially into or out of the tetrahedra and interact ferromagnetically (FM), which is highly analogous to the water-ice rule [20]. According to the common water ice (Bernal-Fowler rules), oxygen atoms sit at the centre of tetrahedra and are surrounded by four hydrogen ions. Two of these four hydrogen ions, are closer to the oxygen at the centre than the other two. Hence, every tetrahedron has six possibilities for achieving this lowest energy state. This leads to residual entropy in a macroscopic state at absolute zero temperature, as described by Pauling for water ice, with a value of $(1/2)Rln(3/2) = 1.68\, J\, mol^{-1} K^{-1}$ [21]. The same analogy of magnetic spin is seen in the pyrochlore lattice (Figure 1). Here, the corner-sharing tetrahedra are associated with two-state spins that approximate classical Ising spins, pointing in a highly degenerate 2-in-2-out state for each tetrahedron at low temperatures, which cannot be lifted by long-range order [21, 22]. Consequently, the frustration of this classical spin ice (CSI) system results in hosting the Coulomb phase and the existence of magnetic monopoles, which have been extensively studied in theory and experiment [23-28]. Magnetisation measurements of $Dy_2Ti_2O_7$ and $Ho_2Ti_2O_7$ indicate strong magnetic anisotropy and effective FM interaction between the nearest-neighbour rare-earth ions, with no evidence of the predicted long-range magnetic ordering down to 60 mK [29-32]. Moreover, the recent focus on quantum spin ice (QSI) observed in pyrochlore materials is gaining attention, as QSI is a special type of quantum spin liquid (QSL). In these materials, stronger quantum fluctuations transform classical spin ice (CSI) into QSI, exploring U(1) gauge theory [3, 33-36]. Interestingly, QSI exhibits emergent quantum electrodynamics with excitations of quantum monopoles and exotic photon-like elementary excitations, as realised in materials such as $Yb_2Ti_2O_7$ [10, 37], $Tb_2Ti_2O_7$ [38], $Ce_2Zr_2O_7$ [39-42], $Ce_2Sn_2O_7$ [8, 39, 43], $Pr_2Hf_2O_7$ [35, 44], $Pr_2Zr_2O_7$ [45, 46], $Pr_2Sn_2O_7$ [47, 48], $Tb_2ScNbO_7$ [49] and $Yb_2GaSbO_7$ [50]. The $Ce^{3+}$ ion (J = 1/2) pyrochlores $Ce_2Sn_2O_7$ and $Ce_2Zr_2O_7$ have been proposed as quantum spin liquid candidates, showing a lack of magnetic order down to at least T = 0.035 K, as evidenced by magnetic, heat capacity, muon spin spectroscopy, and neutron scattering experiments [8, 39, 40, 43]. However, Gd-based pyrochlore compounds ($Gd_2B_2O_7$) with non-magnetic cations in the B site provide a good realisation of antiferromagnetically coupled nearest-neighbour Heisenberg spins on the frustrated pyrochlore lattice. This is because the $Gd^{3+}$ ion possesses a half-filled 4f-shell with nominally no orbital angular momentum (L = 0), resulting in spin-only total angular momentum. As a result, single-ion anisotropy is reduced, and the crystalline electric field effect is negligible. Several gadolinium-based pyrochlores have been synthesised and reported, with most exhibiting an antiferromagnetic ordering transition at T ≤ 1.6 K. Examples include $Gd_2Ti_2O_7$ [51] and $Gd_2Pb_2O_7$ [52], which show antiferromagnetic transitions to long-range magnetic order at $T_N$ = 0.97 K [53, 54] and 0.81 K, respectively [52]. Other compounds include $Gd_2Ge_2O_7$ ($T_N$ = 1.4 K) [55], $Gd_2Pt_2O_7$ ($T_N$ = 1.6 K) [55, 56] and $Gd_2Sn_2O_7$ ($T_N$ = 1 K) [57], all of which display a sharp jump in specific heat anomaly, indicating a strong first-order transition. The magnetic specific heat of $Gd_2(Ge, Pt)_2O_7$ follows the $T^3$ ($C_{mag} \propto T^3$) dependence at low temperatures, corresponding to gapless spin-wave excitations for a conventional three-dimensional antiferromagnet. In contrast, $Gd_2Sn_2O_7$ shows a $T^2$ dependence of $C_{mag}$, indicative of a gapped spin-wave spectrum [58]. Similar compounds such as $Gd_2Zr_2O_7$ and $Gd_2Hf_2O_7$ exhibit sharp peaks around 0.77 K, consistent with the presence of long-range magnetic order [59]. Various classifications of magnetism have been observed across the rare earth pyrochlores, but none of the Gd-based pyrochlore compounds have been reported to exhibit quantum spin ice (QSI) behaviour. However, the garnet lattice $Gd_3Ga_5O_{12}$ (gadolinium gallium garnet, GGG) with corner-sharing triangles of Gd ions has been reported as a quantum spin liquid candidate [60]. Recently, the spin-liquid state in GGG was shown to be associated with long-range hidden order, where multipoles are formed from 10-spin loops in a three-dimensional arrangement of corner-sharing triangles known as hyperkagome due to the interplay between antiferromagnetic spin correlations and local magnetic anisotropy [61]. In contrast, a similar study of gadolinium aluminium garnet (GAG) $Gd_3Al_5O_{12}$[62] found clear ordering transitions at 175 mK [63]. Both GGG and GAG exhibit field-induced antiferromagnetic long-range order when subjected to an applied magnetic field of around 1 T [62, 64].

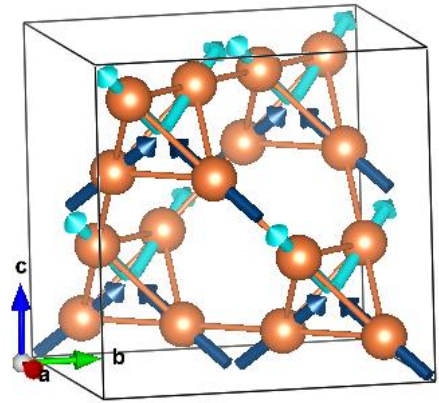

Figure.1. Pyrochlore lattice of corner sharing tetrahedra as occupied by the magnetic atoms (Gd), arrows pointed the magnetic spin moments in spin ice materials.

In this work, we have prepared polycrystalline samples of $Gd_2GaSbO_7$ (GGSO) and $Gd_2InSbO_7$ (GISO), which have not been studied in detail. Previous reports of magnetic measurements for GGSO and GISO indicate no long-range order down to 2 K [65]. Recently, Ortiz et al.[66] reported a family of mixed B-site $Ln_2InSbO_7$ compounds, where magnetic characterisation suggested that this family of compounds may exhibit exotic magnetic ground state properties [67]. Hence, we aim to demonstrate low-energy excitations, such as quantum fluctuations in the spin ice state, in these compounds via low-temperature specific heat and magnetic measurements.

**Experimental methods**

Polycrystalline samples of $Gd_{22}GaSbO_7$ and $Gd_2InSbO_7$ were synthesised using a solid-state reaction method. The analytical mixture had a molar ratio of 2:1:1 (Gd: M: Sb) using high-purity chemicals: $Gd_2O_3$ (99.99%, Alfa), M (M = $Ga_2O_3$ (99.99%, Alfa) and $In_2O_3$ (99.9%, Alfa)), and $Sb_2O_5$ (99.998%, Alfa). The compounds were well-grounded in a mortar and pressed into pellets. The obtained pellets were subjected to heat treatment at various temperatures, specifically 873 K, 1073 K, 1273 K, and 1573 K, for 24 hours in air at ambient pressure. Room temperature powder X-ray diffraction using Cu-Kα radiation was employed to examine the phase purity and stoichiometry of the annealed samples. Magnetic properties and heat capacity were measured using the Quantum Design Physical Property Measurement System (PPMS) from 1.8 K to 300 K with an external magnetic field up to 9 T. Heat capacity was measured using the PPMS relaxation technique down to 400 mK.

## Result and Discussion

## Structural details:

X-ray diffraction analysis revealed that both GGSO and GISO exhibited a similar cubic crystal structure with space group Fd-3m (SG No: 227). Rietveld refinement of the data showed good agreement between observed and calculated intensities, confirming the absence of impurities in the samples (Figure 2). The lattice parameters determined from the refinement were a = 10.267(1) Å for GGSO and a = 10.484(7) Å for GISO. Notably, the lattice parameter decreased for $Gd_2GaSbO_7$ compared to $Gd_2InSbO_7$, which was attributed to the smaller ionic radii of Ga compared to In. Tables 1 and 2 provide detailed atomic coordinates and structural parameters for $Gd_2InSbO_7$ and $Gd_2GaSbO_7$, respectively.

**Table 1:** The structural parameters of $Gd_2InSbO_7$ and $Gd_2GaSbO_7$ derived through Rietveld refinement

| $Gd_2InSbO_7$ | | | | | |
|---|---|---|---|---|---|
| **Atom** | Site | x | y | Z | Occ |
| **Gd** | 16c | 0.00000 | 0.00000 | 0.00000 | 1.00000 |
| **In** | 16d | 0.50000 | 0.50000 | 0.50000 | 0.50006 |
| **O(1)** | 8a | 0.12500 | 0.12500 | 0.12500 | 1.00012 |
| **O(2)** | 48f | 0.40871 | 0.12500 | 0.12500 | 1.00004 |
| **Sb** | 16d | 0.50000 | 0.50000 | 0.50000 | 0.50006 |
| $Gd_2GaSbO_7$ | | | | | |
| **Atom** | Site | x | y | Z | Occ |
| **Gd** | 16c | 0.00000 | 0.00000 | 0.00000 | 1.00000 |
| **Ga** | 16d | 0.50000 | 0.50000 | 0.50000 | 0.55806 |
| **O(1)** | 8a | 0.12500 | 0.12500 | 0.12500 | 0.96390 |
| **O(2)** | 48f | 0.41965 | 0.12500 | 0.12500 | 1.08819 |
| **Sb** | 16d | 0.50000 | 0.50000 | 0.50000 | 0.54307 |

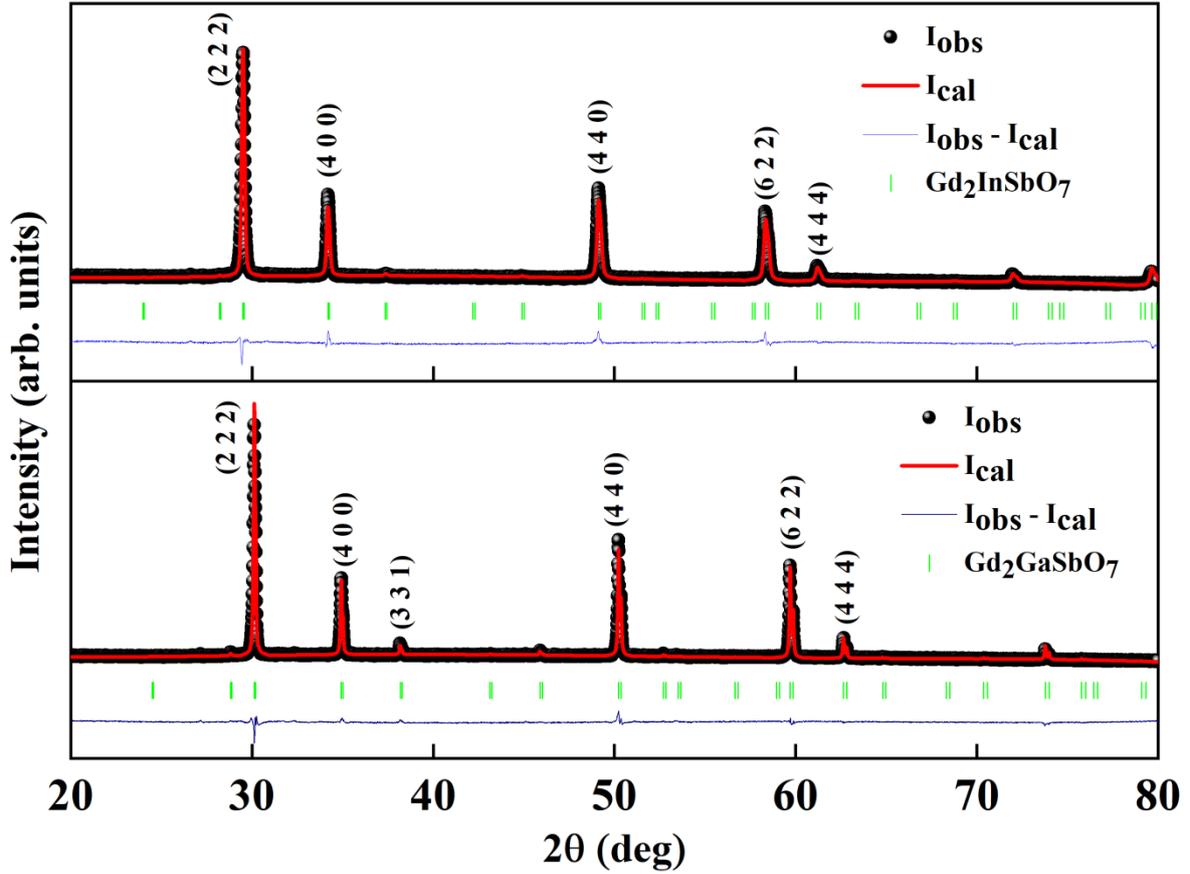

**Figure 2.** Powder X-ray diffraction patterns of GISO and GGSO along with Rietveld refinement using Fullprof program.

## Magnetic susceptibility and magnetisation studies

The temperature dependence of DC magnetic susceptibility at 1 kOe and 10 kOe is shown in Figure 3. These susceptibility measurements do not exhibit a long-range magnetic order down to 2 K, and there is no bifurcation between zero field-cooled (ZFC) and field-cooled (FC) dependences. The temperature dependence of DC susceptibility ($\chi$ = M/H) and inverse susceptibility ($\chi^{-1}$) of GGSO and GISO are shown in the inset of Figures 3a and 3b. The susceptibility does not exhibit any slope change down to 2 K and does not tend to saturate, indicative of the absence of long-range order. Instead, it shows linear behaviour from 2 K to 300 K. The inverse molar susceptibility is consistent with Curie-Weiss behaviour in the temperature range 10–300 K, and fits this temperature range with the Curie-Weiss law.

$$\chi = \frac{C}{(T-\theta_p)} \qquad (1)$$

Where $\theta_p$ is the Curie-Weiss temperature and $C$ is the Curie constant which can be expressed in terms of an effective moment as

$$C = \frac{\mu_{eff}^2 x}{8} \qquad (2)$$

Here $x$ is the number of rare-earth atoms per formula unit. The solid lines in the inset of Figure 3a and 3b are fitted with Equation 1. We obtained the effective magnetic moment, $\mu_{eff(GGSO)}$ = 7.87 $\mu_B$/Gd, and $\mu_{eff(GISO)}$ = 7.82 $\mu_B$/Gd; Curie-Weiss temperature, $\theta_p$ = -9.1 K and $\theta_p$ = -7.1 K; for the compounds GGSO and GISO, respectively. The obtained values of the effective magnetic moment are close to the expected theoretical value for $Gd^{3+}$ free ions (7.94 $\mu_B$) and the negative $\theta_p$ implies the antiferromagnetic interaction. The $Gd^{3+}$ ion has seven unpaired electrons in f orbitals (half-filled f electron shell), resulting in quenched orbital angular momentum ($L = 0$) and isotropic spin-only total angular momentum $J = S = 7/2$. Consequently, crystalline electric field effects are likely negligible, as the anisotropic orbital angular momentum is mainly responsible for splitting the (2J + 1) 8-fold ground state multiplet. Therefore, the nearest-neighbour exchange interactions ($J_{nn}$) can be estimated from Curie-Weiss temperature $\theta_p$ using the mean-field result [68],

$$\frac{J_{nn}}{k_B} = \frac{3\theta_p}{2nS(S+1)} \qquad (3)$$

where $n = 6$ is the number of nearest neighbours. This gives $J_{nn}/k_B$ = -0.144 K and -0.113 K for GGSO and GISO, respectively. However, the high spin moments of these metal ions create relatively large dipolar fields. The strength of the nearest-neighbour dipolar interactions $D_{nn}$ is proportional to the square of the moment

$$\frac{D_{nn}}{k_B} = \frac{5}{3} \frac{\mu_0}{4\pi} \frac{\mu_{eff}^2}{r_{nn}^3}. \qquad (4)$$

where $r_{nn}$ is nearest neighbour's distance.

Isothermal magnetisation as a function of magnetic field for GGSO and GISO is measured at various temperatures such as 2 K, 3 K and 4 K, as shown in Figure 4. The magnetisation reaches approximately $6.64\mu_B/Gd$ for GGSO and $6.71\mu_B/Gd$ for GISO at 2 K, in the presence of an applied magnetic field at 90 kOe, showing a tendency to saturate beyond 90 kOe. However, the observed magnetisation at 90 kOe is slightly smaller than the expected theoretical saturation magnetisation of free ion $Gd^{3+}$ ($M_{sat} = 7\mu_B/Gd$). Also, no coercivity and remanence are observed in the magnetisation versus field (H). For further temperatures 3K and 4K, a similar trend was observed up to 90 kOe, which completely describes the non-interacting magnetic moments in the paramagnetic region. The loss of magnetisation (~5 % at 90 kOe and 2K) indicates that part of the magnetic moment does not align with the magnetic field. This is attributed to the canting of the moments to prevent the long-range dipolar interactions; similar effect was observed in the garnet systems $Gd_3Al_5O_{12}$ and $Gd_3Ga_5O_{12}$[62].

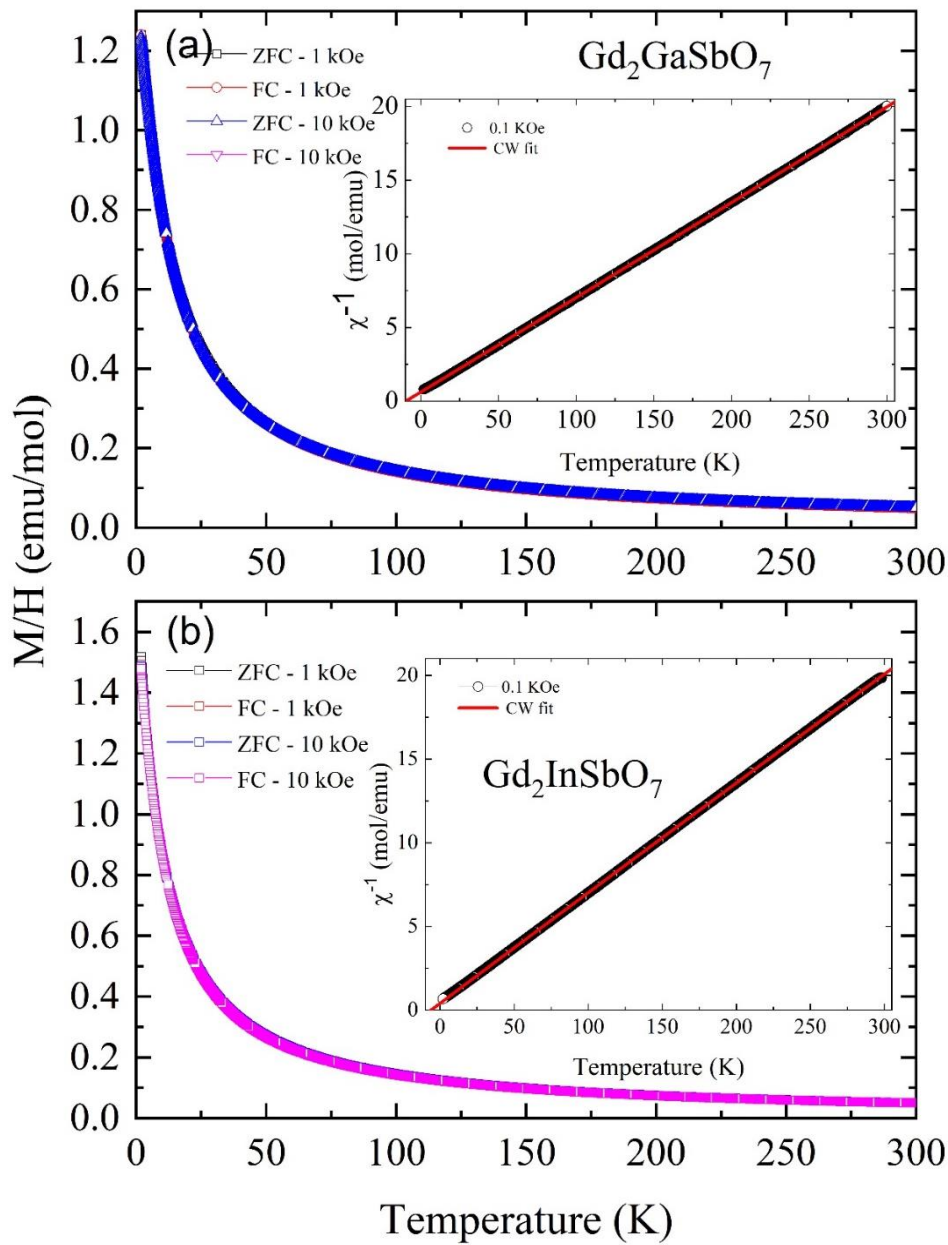

**Figure 3.** Magnetic susceptibility ($\chi$) of GGSO (a) and GISO (b) plotted in the temperature range 2 K – 300 K, measured in the applied magnetic fields of 1 kOe and 10 kOe. Inset: inverse of the magnetic susceptibility ($\chi^{-1}$), with the solid lines representing the fits to the Curie-Weiss (CW) relation.

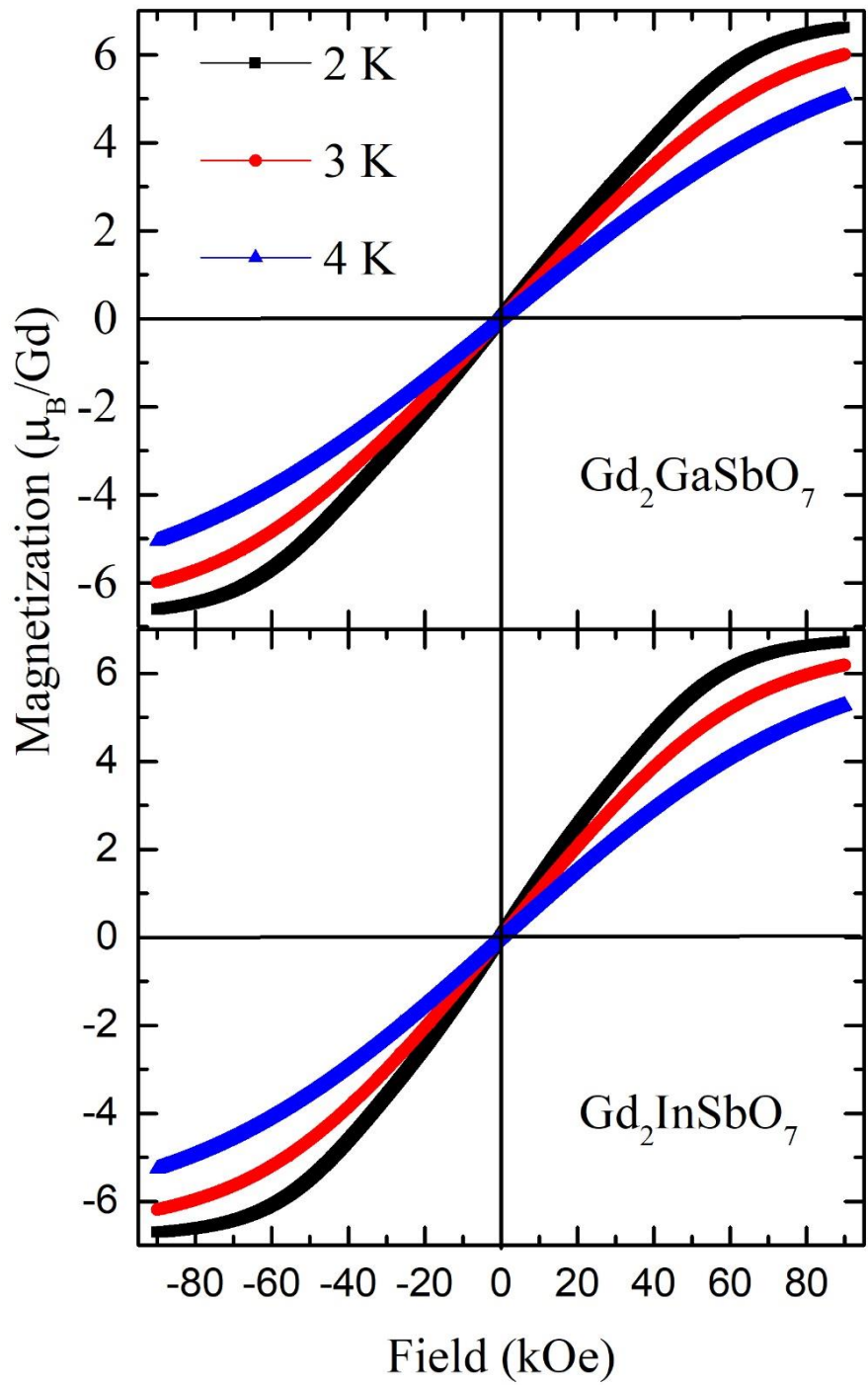

**Figure.4.** Isothermal magnetisation curves of GGSO and GISO as a function of magnetic field up to 90 kOe at selected temperatures.

## Heat Capacity

The temperature dependence of heat capacity measurement was performed down to 400 mK at zero field as shown in Figure 5. We observed a broad maximum around $T_{max}$ = 0.9 K for GGSO and $T_{max}$ = 1 K for GISO, respectively. This indicates short-range magnetic correlations rather than long-range magnetic order, which persists from 15 K down to low temperatures. The magnetic heat capacity ($C_{mag}$) was determined by subtracting the non-magnetic lattice contribution from the total heat capacity. The lattice contribution was estimated by fitting the experimental data with an equation consisting of a linear combination of Debye and Einstein functions.

$$C_V = \gamma T + 9N_A K_B \left(\frac{T}{\theta_D}\right)^3 \int_0^{\frac{\theta_D}{T}} \frac{x^4 e^x}{(e^x-1)^2} dx + 3N_A K_B \left(\frac{T_e}{T}\right)^2 \frac{\exp\left(\frac{T_e}{T}\right)}{\left(\exp\left(\frac{T_e}{T}\right)-1\right)^2} \quad (5)$$

The fitted curve provides reasonable agreement with the experimental data (Figure 5). The obtained parameters are Sommerfeld coefficient $(\gamma) = 135 \; mJ/mol \; K^2$, $\theta_D = 240$ K, and $T_e$ = 428 K for GGSO. For GISO, the parameters are $\gamma = 114 \; mJ/mol \; K^2$, $\theta_D = 256$ K, and $T_e$ = 534 K for GISO, where $\theta_D$ is the Debye temperature and $T_e$ is an Einstein temperature, respectively. The calculated magnetic contribution to the heat capacity, $\Delta C_{mag}$ was observed as $15.73 \; Jmol^{-1}K^{-1}$ and $16.31 \; Jmol^{-1}K^{-1}$ for GGSO and GISO, respectively. This is much smaller than the expected value from mean-field theory, $\Delta C_{mag} = 5R \; S(S+1)/[S^2 + (S+1)^2] = 20.15 \; Jmol^{-1}K^{-1}$ due to high spin state of $Gd^{3+}$ (S = 7/2) for second order magnetic phase transition. This indicates no sharp phase transition like first-order transition in similar Gd-based pyrochlore materials [55]. Furthermore, the magnetic entropy $S_{mag}(T)$ is estimated by integrating the $C_{mag}/T$ and it is plotted in Figure 5.

$$S_{mag} = \int_{0.4}^{T} \frac{C_{mag}}{T} \quad (6)$$

The magnetic entropy $S_{mag}(T)$ saturates around 10 K with values of 30 J $K^{-1}mol^{-1}$ and 30.25 J $K^{-1}mol^{-1}$ for GGSO and GISO, respectively. These values are smaller than the expected value from the mean field theory for an $S = 7/2$ spin system $\Delta S_{mag} = 2Rln(2S+1) = 34.57$ J $K^{-1}mol^{-1}$. where R is the ideal gas constant. Pauling's prediction for the entropy of water ice, $2R[ln(8) - (1/2) ln(3/2)] = 31.21 \; Jmol^{-1}K^{-1}$ [21], where the factor of 2 accounts for the number of magnetic atoms in the unit cell. provides good agreement with the residual entropy observed in these compounds due to geometrical frustration in Gd-based pyrochlore, as shown in Figure 5.

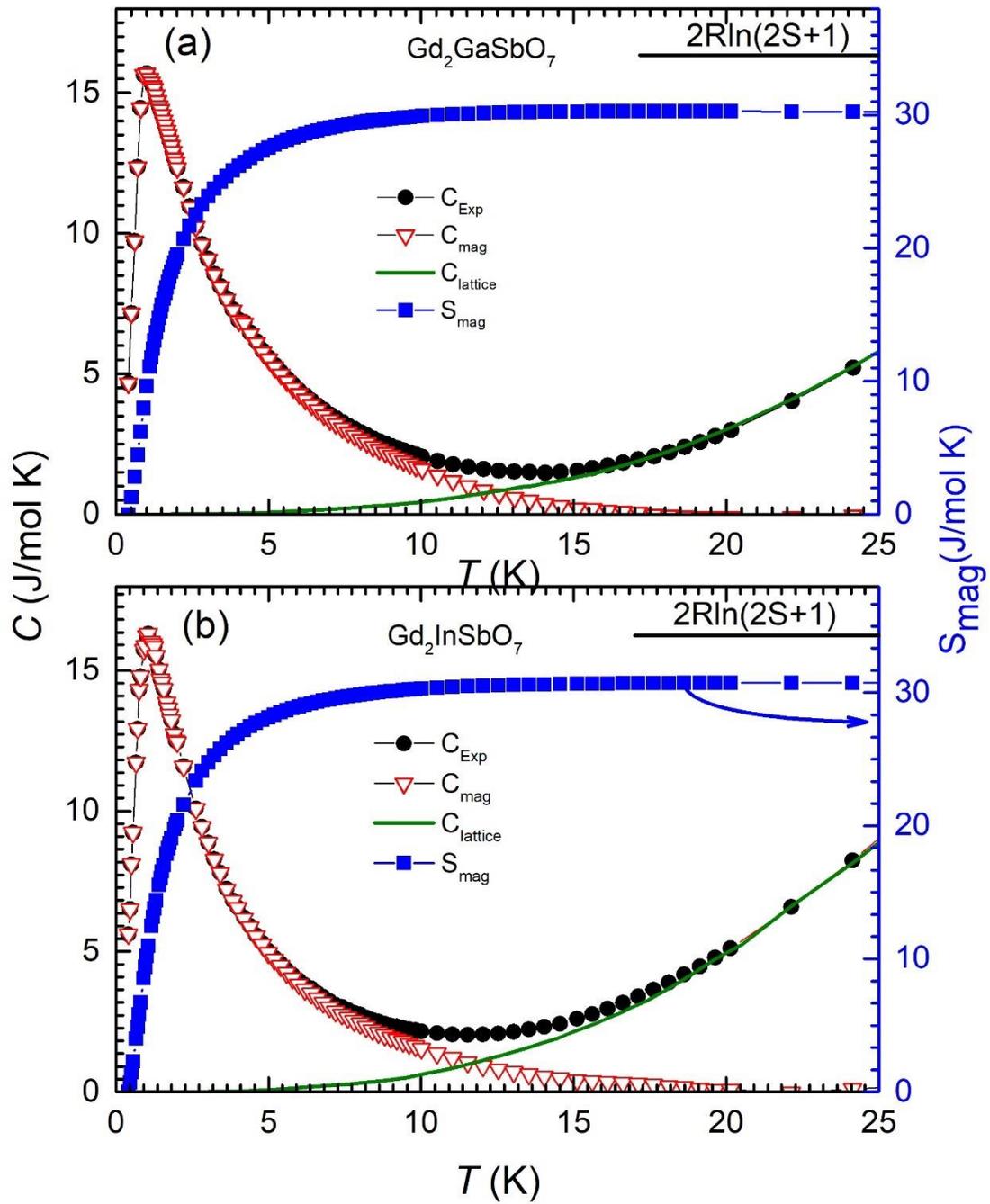

**Figure.5.** Temperature-dependent specific heat of GGSO and GISO measured from 400 mK up to higher temperatures. The data is fitted with a linear combination of Debye and Einstein functions. The figure also includes plots of the magnetic heat capacity ($C_{mag}$) and magnetic entropy ($S_{mag}$) as functions of temperature, alongside the comparison with the total entropy Rln8 in the systems.

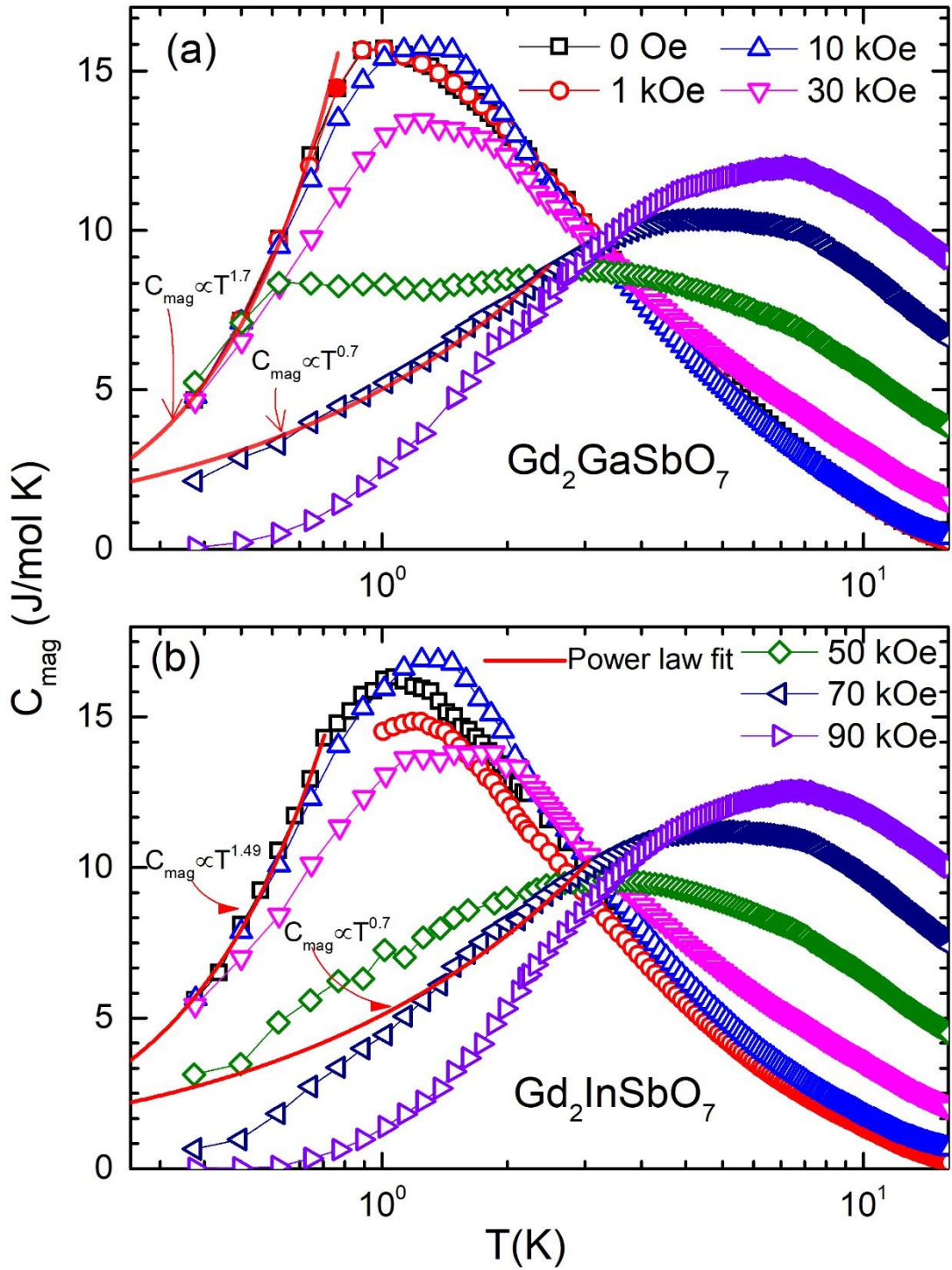

Figure 6. Low-temperature heat capacity ($C_{mag}$) of polycrystalline GGSO and GISO measured in zero fields and under applied magnetic fields up to 90 kOe. The solid line represents the powder law fit for $C_{mag} \approx T^{d/n}$ behaviour.

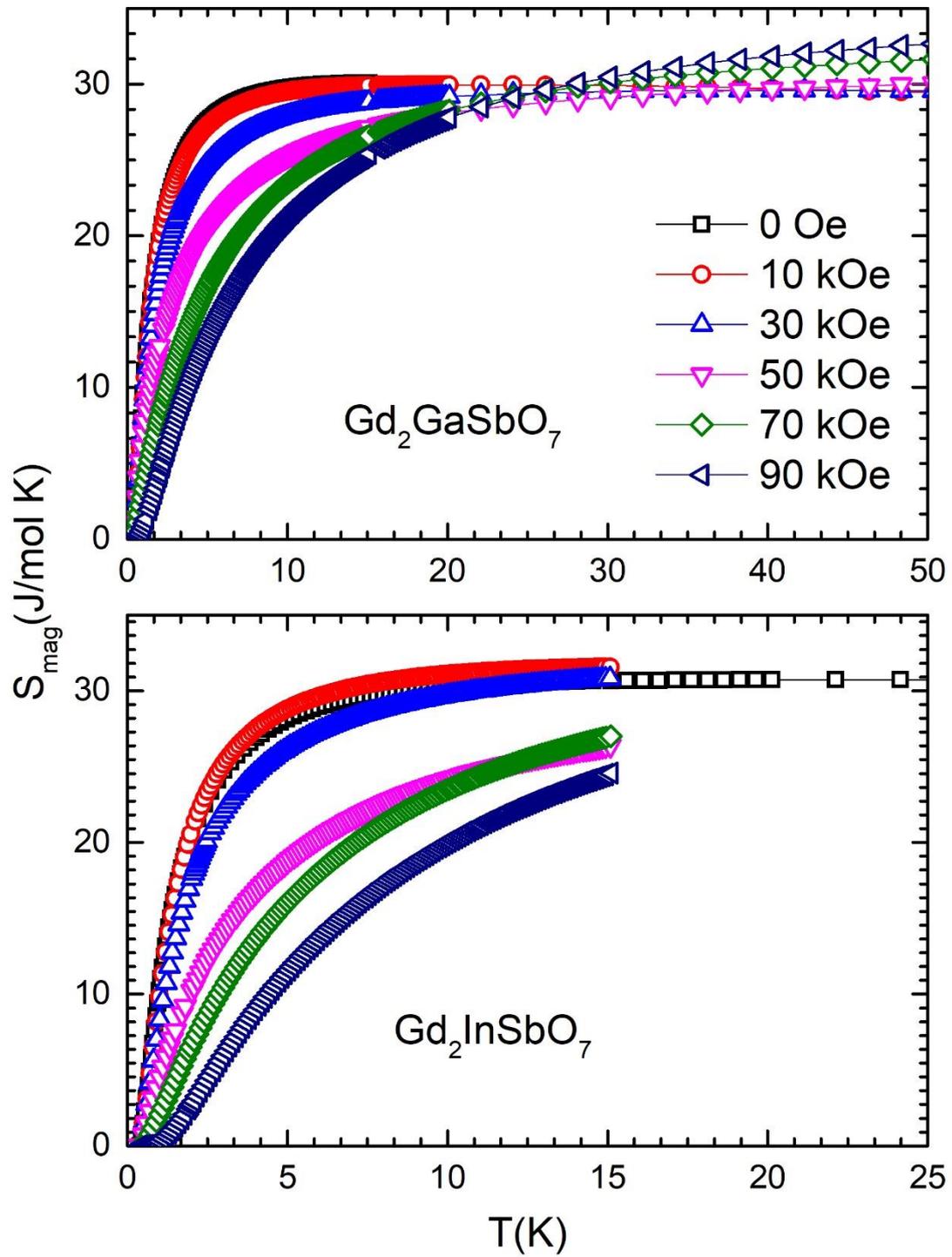

Figure 7. Change in magnetic entropy ($S_{mag}$) versus temperature measured under various applied magnetic fields ranging from 0 to 90 kOe for polycrystalline GGSO and GISO.

The analysis of magnetic specific heat ($C_{mag}$) below ordering temperature in terms of spin wave (SW) theory gives $C_{mag} \approx T^{d/n}$ for magnons, where d is the dimensionality constant for magnetic lattice, and n is the dispersion relation $\omega \approx \kappa^n$. For a conventional 3D antiferromagnetic insulator, $C_{mag}$ is directly proportional to $T^3$ below the ordering temperature ($T_N$). When fitting the power law below the $T_{max}$, it exhibits the non-integer values of $d/n$ = 1.71 and 1.49 for GGSO and GISO, respectively, as shown in Figure 6. Due to this non-integer $T^3$ power law fitting, the spin wave spectrum cannot be approximated by a gapped linear dispersive mode like $Gd_2Sn_2O_7$ [57]. Additionally, the frustrated spin disorder can produce non-$T^3$ behaviour in specific heat at low temperatures, similar to what has been reported in 3D quantum spin liquid candidates like $PbCuTe_2O_6$ [69, 70]. As seen in Figure 6. for the magnetic contribution in specific heat, the magnetic field suppresses the $T_{max}$ under a magnetic field below 50 kOe. Then, it is shifted to higher temperatures along with an enhancement in the height of $T_{max}$ in $C_{mag}$. This unusual behaviour, such as broad maximum and enhancement in height with increasing magnetic field in $C_{mag}$ was also observed in reported quantum spin liquid compounds like $Ce_2Zr_2O_7$, $Yb_2GaSbO_7$ and $YbMgGaO_4$ [39, 50, 71]. Furthermore, the magnetic entropy $S_{mag}$ also increases with higher magnetic fields as shown in Figure 7. Such behaviour has been found in the CSI and QSI compounds [12, 21, 35, 48, 50, 72, 73].

## Discussion

Generally, rare-earth-based pyrochlore compounds are used to investigate geometrically frustrated magnetism [52, 74-78]. $Gd^{3+}$ has a high spin of $S$ =7/2 moments; hence, Gd-based pyrochlore compounds should be a good representation of the classical Heisenberg antiferromagnet, as summarised in Table 2. All the reported Gd-based pyrochlores exhibited long-range magnetic order below 1.5 K. In this study, due to $Gd^{3+}$ ions, the compounds GGSO and GISO are expected to be explained by the same classical Heisenberg antiferromagnetic interaction[55]. However, our results show the absence of long-range order in heat capacity and magnetic measurements, which is entirely different from all other isostructural compounds reported so far (shown in Table 2). To investigate this interesting phenomenon, we calculated the nearest neighbour exchange interaction ($J_{nn}/k_B$) from $\theta_P$ as $J_{nn}/k_B$ = -0.144 K for GGSO and $J_{nn}/k_B$ = -0.113 K for GISO, respectively. These values are compared with other analogies of Gd-pyrochlore compounds[51, 53, 55-58, 79]. The GISO has the lowest values among these compounds, indicating that the next-nearest neighbour interactions might be weaker. Since the lattice parameters have small variations depending on the ionic radius for other isostructural analogies of this series, the calculated dipolar interaction ($D_{nn}/k_B$) is dependent on effective magnetic moments and interatomic distance between the magnetic ions. The obtained values are $D_{nn}/k_B$ = 1.346 K for GGSO and $D_{nn}/k_B$ = 1.247 K for GISO, respectively. These two compounds have intermediate values among all other $Gd^{3+}$ compounds. However, dipole-dipole interaction ($D_{nn}/k_B$) is more dominant in these pyrochlore compounds than the nearest neighbours interaction ($J_{nn}/k_B$). Therefore, this series of materials is characterised by the ratio $J_{nn}/D_{nn}$ for further analysis, as shown in Figure 8(a). It is almost greater than 0.09 for all the Gd-based compounds [78, 80]. GISO has the lowest ratio of $J_{nn}/D_{nn}$, indicating that weaker Heisenberg interaction could cause frustration[72].

Additionally, a broad feature was observed in heat capacity centred around ~1 K, which is more prominent in GISO than in GGSO. This broad feature is attributed to the build-up of spin fluctuations at low temperatures. This feature in heat capacity shifts to higher temperatures with an applied magnetic field over the range from 0 to 90 kOe (Figure 6). Moreover, $C_{mag}$ follows the exponential function ($C_{mag} \propto T^{-2}exp(-\Delta/T)$) from 400 mK to 1 K. Possibly due to gapped spin wave excitation at low temperatures [57]. Hence, we plotted $log(C_{mag}T^2)$ versus $1/T$ as shown in Figure 8(b), where the linear fit gives the $\Delta \approx 0.9\,K$ for both GGSO and GISO. The gap might be attributed to the anisotropy and long-range dipolar interaction [55, 63]. However, it is recommended to measure below 400 mK for a clear picture of ground state excitation.

The important finding of this paper is that GGSO and GISO undergo quantum fluctuations in spin ice-like systems[73], as indicated by specific heat measurements showing a broad maximum of around 1 K and residual entropy akin to that of water ice. The major differences in the series of Gd-based pyrochlores are the Gd-Gd distance ($r_{nn} = a/4\sqrt{2}$, where $a$ is lattice parameter) and cell density due to changes in the B-site ions (shown in Figure 8(c)). In GGSO and GISO compounds, the B-site is randomly occupied by a 1:1 ratio with $Ga^{3+}$: $Sb^{5+}$ and $In^{3+}$: $Sb^{5+}$ with different ionic radii and electronic configurations. Thus, the enhancement of geometrical frustration in these compounds might be due to superexchange interactions mediated through B site cations like Gd-O-Ga-O-Gd, Gd-O-In-O-Gd and Gd-O-Sb-O-Gd. Similar features were observed in $Yb_2GaSbO_7$ [50] and $Ln_2InSbO_7$(Ln is a rare earth element)[66]. Another possibility that may cause the magnetic frustration effect in these GGSO and GISO compounds is the potential chemical disorder of $Gd^{3+}$ ions on $Ga^{3+}$ and $In^{3+}$ sites. However, the ionic radius of $Gd^{3+}$ is larger than that of $Ga^{3+}$ and $In^{3+}$ with the same valence charges. Structural characterisation from XRD Rietveld refinement is not sufficient to reveal this ionic disorder. Furthermore, the negative $\theta_p$ indicates antiferromagnetic correlations, which differ from the ferromagnetic interactions exhibited by classical spin ice compounds [72]. However, the QSI state exhibited in the pyrochlore structure with antiferromagnetic interaction (it is U(1) QSL)[34-36], even in this QSI regime, some materials conserve residual entropy similar to classical spin ice, such as $Pr_2Zr_2O_7$[73], $Pr_2Sn_2O_7$[48] and $Yb_2GaSbO_7$[50]. GGSO and GISO also mimic this behaviour. Notably, Ortiz et al.[66] reported a low-temperature (down to T = 350 mK) ac and dc magnetic susceptibility for GISO, where $T_{max}$ exhibits no frequency dependence of ac susceptibility, clearly indicating the absence of spin glass behaviour in GISO. Therefore, the presence of residual entropy and the absence of a sharp anomaly in heat capacity at zero applied magnetic fields suggest the possibility of a dynamical ground state property in GGSO and GISO. Further neutron diffraction and muon spin relaxation techniques are needed better to understand these materials' magnetic ground state properties. Additionally, studies on single crystals of GGSO and GISO could explore the mechanism of frustration in these compounds.

**Table 2:** The structural and magnetic properties of $Gd_2GaSbO_7$ and $Gd_2InSbO_7$ are compared with other Gd-based pyrochlore compounds. $\theta_P$: Curie-Weiss temperature; $T_{mag}$ (K): Magnetic transition temperature; $\mu_{eff}$ ($\mu_B$): Effective magnetic moment of rare earth; $a$: Lattice parameter; $J_{nn}/k_B$ (K): The exchange energy for the two nearest neighbour ions; $D_{nn}/k_B$ (K): The dipole-dipole interaction energy between the nearest neighbour ions at low temperatures.

| Compound | $\theta_P$ (K) | $T_{mag}$ (K) | $\mu_{eff}$ ($\mu_B$) | $a$ (Å) | $J_{nn}/k_B$ (K) | $D_{nn}/k_B$ (K) |
|---|---|---|---|---|---|---|
| $Gd_2Pt_2O_7$[56] | -9.4 | 1.6 | 7.83 | 10.2626 | -0.149 | 1.333 |
| $Gd_2Sn_2O_7$[55, 56] | -8.6 | 1 | 8.06 | 10.454 | -0.136 | 1.337 |
| $Gd_2Zr_2O_7$[59] | -7.7 | 0.769 | 7.83 | 10.522 | -0.122 | 1.238 |
| $Gd_2Hf_2O_7$[59] | -7.3 | 0.771 | 7.80 | 10.489 | -0.116 | 1.154 |
| $Gd_2Ti_2O_7$[55] | -8.8 | 0.871 | 7.80 | 10.182 | -0.139 | 1.355 |
| $Gd_2Ge_2O_7$[55] | -11.1 | 1.40 | 8.05 | 9.9985 | -0.176 | 1.524 |
| $Gd_2Pb_2O_7$[[52] | -7.35 | 0.8 | 7.67 | 10.7292 | -0.117 | 1.120 |
| $Gd_2Ir_2O_7$[14, 79] | -7.9 | - | 7.91 | 10.290 | -0.125 | 1.350 |
| $Gd_2GaSbO_7$* | -9.1 | 0.91 | 7.87 | 10.2671 | -0.144 | 1.346 |
| $Gd_2InSbO_7$* | -7.1 | 1.06 | 7.82 | 10.4847 | -0.113 | 1.247 |

*This work

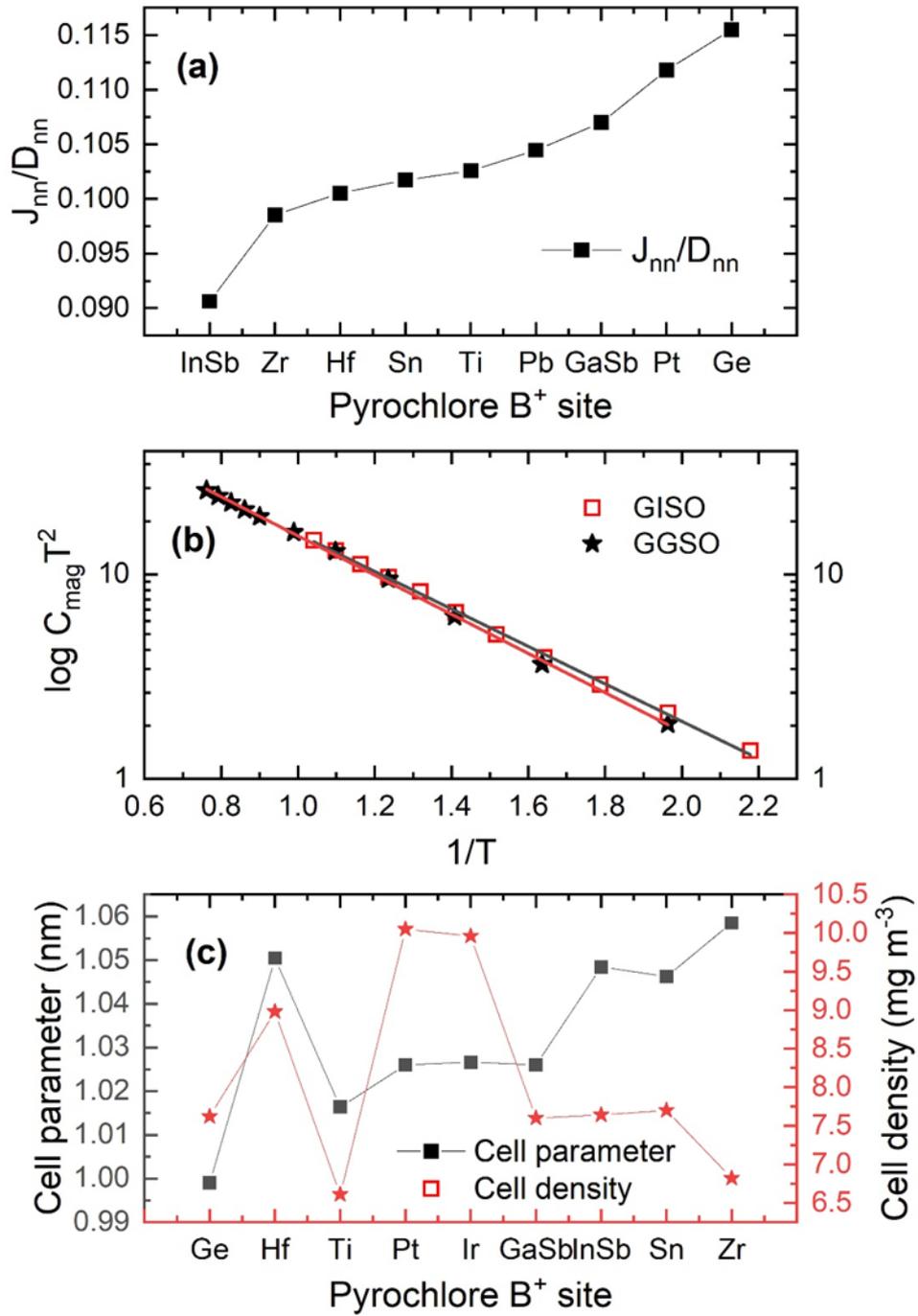

Figure 8. (a). Different Gd-based pyrochlores with the ratio between nearest neighbour exchange ($J_{nn}$) and dipolar interaction energy ($D_{nn}$). (b). log($C_{mag}T^2$) plotted versus 1/T at low temperature limit from 400 mK to 1.5 K, Where $C_{mag} \propto T^{-2}exp(-\Delta/T)$ follow the straight line. (c). Cubic lattice parameter versus cell density with different Gd-based pyrochlore compounds.

## Conclusions

In summary, we have observed evidence of a dynamical ground state in spin ice-like polycrystalline GGSO and GISO pyrochlore systems. These compounds were synthesized, and magnetic measurements were conducted down to 2 K, and specific heat measurements down to 400 mK, under an applied magnetic field of up to 90 kOe. A broad maximum centred around 1 K was observed in the heat capacity measurements, indicating the absence of long-range magnetic order in these compounds. Both GGSO and GISO retain residual entropy similar to that of water ice, confirming the presence of a spin ice-like state. Negative Curie-Weiss temperature ($\theta_p$ = -9.1 K for GGSO and $\theta_p$ = -7.1 K for GISO); indicate antiferromagnetic interactions. This behaviour contrasts with classical spin ice compounds such as $Dy_2Ti_2O_7$ and $Ho_2Ti_2O_7$. A comparison of all reported Gd pyrochlore compounds reveals that GISO has the lowest ratio of nearest neighbour exchange interaction to dipolar interactions $\left(\frac{J_{nn}}{D_{nn}} \approx 0.09\right)$. This suggests that different ionic pathways of exchange interaction may influence the absence of long-range magnetic order. At low temperatures, magnetic heat capacity followed a $T^{1.71}$ and $T^{1.49}$ dependence for GGSO and GISO, respectively, indicating gapped magnetic excitations. This behaviour could provide insights into the realization of magnetic monopoles in a condensed state. However, further studies using neutron diffraction and muon spin relaxation are needed to better understand the magnetic ground state. The exotic spin dynamics observed in these compounds offer new perspectives for understanding geometric frustration in magnetic systems.

## Acknowledgement

This work results from implementing the following projects: University Science Park TECHNICOM for Innovation Applications Supported by Knowledge Technology – II-P Phase, ITMS: 313011D232, supported by the Research & Development Operational Programme funded by the ERDF and also by projects VEGA 1/0180/23; VEGA 1/0407/24. The author S.N thanks VIT-AP university, India for supporting this work under the RGEMS project (No: VIT-AP/SPORIC/RGEMS/2022-23/009).